\documentclass[american,aps,preprint]{revtex4-1}
\usepackage{graphicx}
\usepackage{bm}
\usepackage{color}

\usepackage[american]{circuitikz}

\usepackage[T1]{fontenc}
\usepackage[utf8]{inputenc}
\usepackage{babel}
\usepackage{array}
\usepackage{booktabs}
\usepackage{textcomp} 
\usepackage{physics}
\usepackage{lineno}

\usepackage{latexsym,amsmath,amssymb,mathtools}

\usepackage[unicode=true,
bookmarks=true,bookmarksnumbered=false,bookmarksopen=false,
breaklinks=false,pdfborder={0 0 0},pdfborderstyle={},backref=false,colorlinks=true]
{hyperref}
\hypersetup{
	colorlinks=true,linkcolor=blue,citecolor=blue}

\usepackage[most]{tcolorbox}
\definecolor{myblue}{rgb}{.8, .8, 1}
\usepackage{empheq}

\newlength\mytemplen
\newsavebox\mytempbox

\makeatletter
\newcommand\mybluebox{%
	\@ifnextchar[%]
	{\@mybluebox}%
	{\@mybluebox[0pt]}}
\def\@mybluebox[#1]{%
	\@ifnextchar[%]
	{\@@mybluebox[#1]}%
	{\@@mybluebox[#1][0pt]}}
\def\@@mybluebox[#1][#2]#3{
	\sbox\mytempbox{#3}%
	\mytemplen\ht\mytempbox
	\advance\mytemplen #1\relax
	\ht\mytempbox\mytemplen
	\mytemplen\dp\mytempbox
	\advance\mytemplen #2\relax
	\dp\mytempbox\mytemplen
	\colorbox{myblue}{\hspace{1em}\usebox{\mytempbox}\hspace{1em}}}

\newtcbox{\mymath}[1][]{%
	nobeforeafter, math upper, tcbox raise base,
	enhanced, colframe=blue!30!black,
	colback=blue!30, boxrule=1pt,
	#1}
\usepackage{xcolor}
\definecolor{lightgreen}{HTML}{90EE90}

\graphicspath{{figures/}}

\begin{document}
\title{Self-organization in the one-dimensional Landau-Lifshitz-Gilbert-Slonczewski equation with non-uniform anisotropy fields}
\author{M\'onica A. Garc\'ia-\~Nustes}
\email{monica.garcia@pucv.cl }
\affiliation{Instituto de F\'isica, Pontificia Universidad Cat\'olica de Valpara\'iso, Casilla 4059, Chile}
\author{Fernando R. Humire}
\email{f.humire@academicos.uta.cl}
\affiliation{Departamento de F\'isica, Facultad de Ciencias, Universidad de Tarapac\'a, Casilla 7-D Arica, Chile}

\author{Alejandro O. Leon}
\email{alejandroleonvega@gmail.com }
\affiliation{Center for the Development of Nanoscience and Nanotechnology, CEDENNA, Santiago, Chile}

\begin{abstract}	
In magnetic films driven by spin-polarized currents, the perpendicular-to-plane anisotropy is equivalent to breaking the time translation symmetry, i.e., to a parametric pumping. In this work, we numerically study those current-driven magnets via the Landau-Lifshitz-Gilbert-Slonczewski equation in one spatial dimension. We consider a space-dependent anisotropy field in the parametric-like regime. The anisotropy profile is antisymmetric to the middle point of the system. We find several dissipative states and dynamical behavior and focus on localized patterns that undergo oscillatory and phase instabilities. Using numerical simulations, we characterize the localized states' bifurcations and present the corresponding diagram of phases.
\end{abstract}
\maketitle

%%%%%%%%%%%%%%%%%%%%%%%%%%%%%%%%%%%%%%%%%%%%%%%%%%%%%%%%%%%%%%%%%%%%%%%%%%%%%%%%%%%%%%%%%%%%%%%%%%%%%%%%%%%%%%%%%%%%%%%%%%
\section{Introduction}
Non-equilibrium systems present complex dynamics~\cite{LibroStrogatz,Cross:1993tk}, including pattern formation~\cite{Cross:1993tk,Hoyle2006pattern}, localized states~\cite{Cross:1993tk}, and chaotic behaviors~\cite{LibroStrogatz}. 
Nanometric magnetic systems exhibit quasi-Hamiltonian dynamics, perturbed by a relatively small injection and dissipation of energy~\cite{Mayergoyz2009}. Domain walls~\cite{Stiles2006spin,Fronts1}, self-sustained oscillations~\cite{Kiselev,Slavin}, textures~\cite{Volkov:2013bc,Kravchuk:2013dw,WithSaliya,Equivalence}, topological~\cite{Skyrmions,Vortices} and non-topological~\cite{Mikeska,Equivalence,Pulse1,Breather} localized states are examples of non-equilibrium states of driven nanomagnets. These are solutions of a nonlinear partial differential equation that governs the magnetization dynamics in the continuum limit, namely the Landau-Lifshitz-Gilbert-Slonczewski (LLGS) model~\cite{Mayergoyz2009}. Even though Landau and Lifshitz published the first form of this equation in 1935~\cite{LLEqOriginal}, it is still widely investigated and revised to incorporate recently discovered effects. For example, electric currents with a polarized spin~\cite{Slon,Berger} are modeled as a non-variational term that injects energy and favors limit cycles~\cite{Kiselev,Slavin}. Also, the dispersion of magnetization waves can be generalized to include anisotropic terms that create topological textures~\cite{Skyrmions}. A third relevant mechanism is the modulation of the magnetic anisotropy fields by applied voltages in insulating structures. Since these fields are responsible for the saturation of the magnetization near equilibria (i.e., they are the most relevant nonlinearities of the LLGS equation), their tuning by electric fields promises a space or time-dependent control of both the linear and nonlinear parts of the LLGS system.

The voltage-controlled magnetic anisotropy (VCMA) effect~\cite{Suzuki2011,Shiota2009,Kanai2012,Nozaki2012} promises memory devices with low power consumption due to the absence of Joule dissipation. Furthermore, VCMA can induce several magnetization responses. For example, a voltage pulse can assist~\cite{Shiota2009} or produce~\cite{Kanai2012} the switching of the magnetization between two equilibria. Voltages oscillating near the natural frequency generate resonances~\cite{Nozaki2012}. On the other hand, if the voltage oscillates at the twice natural frequency, parametric instabilities~\cite{ParametricVCMA4,ParametricVCMA2}, Faraday-type waves, and localized structures emerge~\cite{Breather}. These phenomena also appear in other \textit{parametrically driven systems} that are excited by a force that simultaneously depends on time and the state variable. Magnetic anisotropies can be manipulated by applying a strain to the magnetic medium, via the modulation of its thickness and doping with heavy atoms. While temporally modulated magnetic anisotropies receive considerable attention, the self-organization arising from its static non-uniform counterpart has not been fully explored.

In this work, we study the Landau-Lifshitz-Gilbert-Slonczewski equation in one spatial dimension, representing a magnet subject to a magnetic field, a spin-polarized charge current~\cite{Kiselev,Slavin,Slon,Berger}, and a non-uniform perpendicular magnetic anisotropy. We focus on the parameter values where the current acts as additional damping and stabilizes an equilibrium that would otherwise be unstable. In this regime, the magnet behaves as a parametrically-driven system, even if its parameters are time-independent. The parametric nature of the magnet manifests as a breaking of symmetry induced by the magnetic anisotropy field, in the same way as a time-varying force breaks the time-translation invariance in parametrically driven systems. Furthermore, the LLGS equation can be mapped to the parametrically driven damped Nonlinear Schr\"odinger equation (pdNLS), which is the paradigmatic model of parametrically forced systems. In this transformation, the current acts as damping, the applied magnetic field is a frequency shift or detuning, and the anisotropy field is equivalent to a parametric injection of energy.
Regarding the anisotropy field, the considered field profile is the sum of two Gaussian functions with opposite signs. We find that localized patterns emerge when the anisotropy field is large enough, as occurs in parametrically driven systems with a heterogeneous excitation~\cite{FaradayHeterogeneo}.  
 
Those localized patterns can be dynamic or stationary, depending on the parameter values. For example, when the absolute value of the anisotropy at the left and the right are not the same, localized patterns drift. Via the calculation of the eigenvalues of the fixed pattern near the drifting transition, we find that a stationary instability is responsible for this bifurcation. This instability is of a subcritical type, which creates bistability between drifting and pinned patterns. If the modulus of the left and right anisotropy fields are similar, an oscillatory (Andronov-Hopf) instability occurs when the modulus of the charge current is below a threshold. Similar results replicate as the separation distance between the left and the right of the anisotropy varies.

The anisotropy profile considered here resembles some properties of the so-called \textit{parity-time} ($\mathcal{PT}$)-\textit{symmetric} systems, which are characterized by internal gain and loss but conserve the total energy. Recently, the generalization of $\mathcal{PT}$-\textit{symmetric} systems to include injection and dissipation of energy as a small perturbation, i.e., \textit{quasi-$\mathcal{PT}$-symmetric systems}, have gathered some attention, see~\cite{Rabi} and references therein. Phase transitions in out-equilibrium magnetic systems, via the breaking of a $\mathcal{PT}$-symmetry, have been reported~\cite{Galda1, Galda2, Barashenkov}. For example, in Refs.~\cite{Galda1, Galda2}, with the use of a spin Hamiltonian with an imaginary magnetic field, one arrives at the LLGS equation~\cite{Galda1}. A phase transition between conservative and non-conservative spin dynamics is discussed in terms of the $\mathcal{PT}$-symmetry and extended to spin chains~\cite{Galda2}. Stable solitons in nearly $\mathcal{PT}$-symmetric ferromagnet with the spin-torque oscillator with small dissipation~\cite{Barashenkov}.

The article is organized as follows. In the next section, we introduce the LLGS equation and transform it into an amplitude equation with broken phase invariance (i.e., the pdNLS equation). In Sec.~\ref{NumericalResults}, we show our numerical results, while our conclusions and remarks are in Sec.~\ref{Conclusions}.

\section{Landau-Lifshitz-Gilbert-Slonczewski equation}
\begin{figure}[b!]
  \begin{minipage}[c]{0.3\textwidth}
    \includegraphics[width=\textwidth]{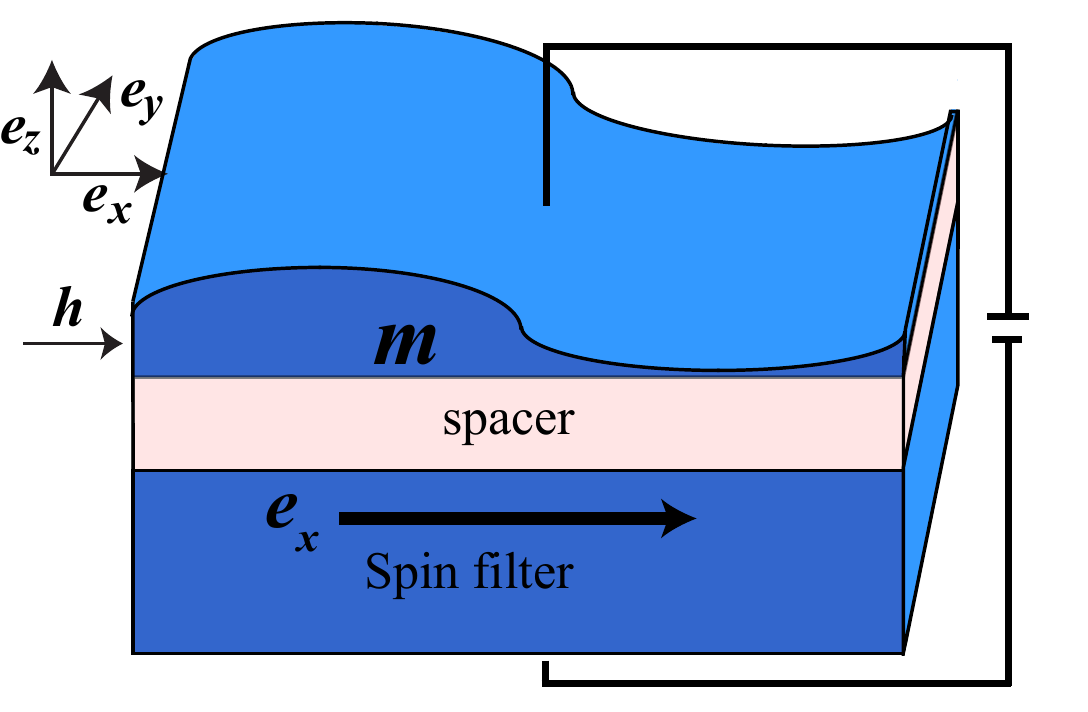}
  \end{minipage}\hfill
  \begin{minipage}[c]{0.6\textwidth}
    \caption{Schematic representation of the setup under study. A magnet with unit magnetization $\mathbf{m}$ is affected by a magnetic field $\mathbf{h}$ and charge current. The current is spin-polarized after traversing a thicker magnet with magnetization along the unit vector $\mathbf{e_x}$. We focus on the layer with a variable perpendicular-anisotropy coefficient, which is represented as a thin layer with variable thickness. } \label{Fig1}
  \end{minipage}
\end{figure}

Let us consider a ferromagnetic medium with two transverse lengths small enough to guarantee that the magnetization remains uniform in those directions (i.e., the magnetization varies along one axis only). Its magnetization is $\mathbf{M}(x,t)=M_s\mathbf{m}(x,t)$, where $M_s$ is the norm, and $\mathbf{m}$ is the unit vector along the orientation of $\mathbf{M}$. The dimensionless space and time coordinates are $x\in [0,L]$ and $t\in [t_0,t_f]$, respectively, where $L$ is the length of the magnet, and $t_0$ and $t_f$ are the initial and end time of the simulation. The magnet is part of a so-called nano-pillar structure, see Fig~\ref{Fig1}(a). There is an applied magnetic field $\mathbf{h}=h\mathbf{e_x}$, where $\{\mathbf{e_x},\mathbf{e_y},\mathbf{e_z}\}$ are the unit vectors along the corresponding Cartesian axis. The spatiotemporal dynamics of the magnetization obeys the Landau-Lifshitz-Gilbert-Slonczewski equation, which in its dimensionless form reads~\cite{Mayergoyz2009}
\begin{align}
\partial_t\mathbf{m}=&-\mathbf{m}
\times\mathbf{h}_\text{eff}
+g\mathbf{m}\times\left(\mathbf{m}
\times\mathbf{e_x}\right)+\alpha\mathbf{m}
\times\partial_t\mathbf{m},
\label{Eq-LLG}\\
\mathbf{h}_\text{eff}=&h\mathbf{e}_{x}+
\partial_{xx}^2\mathbf{m}-h_d(x) m_{z}\mathbf{e}_{z},
\label{Eq-EffField}
\end{align}
where the first term of Eq.~(\ref{Eq-LLG}) induces energy-conservative precessions around the effective magnetic field $\mathbf{h}_\text{eff}$. It has contributions from the external field $h\mathbf{e_x}$, the exchange (or dispersion) field in one spatial dimension $\partial_{xx}^2\mathbf{m}$, and the anisotropy field $-h_d(x) m_{z}\mathbf{e}_{z}$.  The function $h_d$ is the sum of the magnetocrystalline anisotropy and the demagnetizing field in the local approximation. The effective field is the functional derivative of the magnetic energy $E_M$, $\mathbf{h}_\text{eff}\equiv -\delta E_M/\delta\mathbf{m}$, where
\begin{equation}
E_M=\int_0^Ldx\left[-hm_x+\frac{h_d(x)}{2}m_z^2+\frac{\vert \partial_x\mathbf{m}\vert^2}{2}\right].
\end{equation}
The second term of the LLGS equation is the spin-transfer torque~\cite{Slon,Berger}
 induced by the current and parametrized by $g$. This is a non-variational effect that can inject into (for $g>0$) and dissipate (for $g<0$) the magnetic energy. Finally, the third term of Eq.~(\ref{Eq-LLG}) is a phenomenological Rayleigh-like dissipation, ruled by the dimensionless parameter $\alpha$. Typical values~\cite{Mayergoyz2009} of $g$ and $\alpha$ are $g\sim10^{-3}$ and $\alpha\sim10^{-3}-10^{-2}$, respectively.
 
Two equilibria of Eq.~(\ref{Eq-LLG}) exist for all parameter values. In the first equilibrium, the magnetization is parallel to the spin polarization of the charge current ($\mathbf{m}=\mathbf{e_x}$) and the other where they are antiparallel ($\mathbf{m}=-\mathbf{e_x}$). Positive (negative) values of $g$ tend to destabilize (stabilize) the equilibrium $\mathbf{m}=\mathbf{e_x}$. On the other hand, positive (negative) fields $h$ favor (disfavor) the state $\mathbf{m}=\mathbf{e_x}$. Then, in this type of nanometric magnets, there can be two competing effects, the current and the field, and non-trivial non-linear dynamics emerge.

The magnetization dynamics induced by the combination of currents and fields have several similarities~\cite{Equivalence} with systems subject to a forcing that oscillates at twice their natural frequency, namely, \textit{parametrically driven systems}. This association becomes evident when using the so-called Stereographic mapping $ \psi=(m_y+im_z)/(1+m_x)$, which projects the magnetization field on the unit sphere $\mathbf{m}^2=1$ to the complex plane (see Ref.~\cite{Equivalence} and references therein). The resulting amplitude equation reads

\begin{align}
\left(i+\alpha\right){\partial_t} \psi=\left(ig-h\right) \psi-\frac{h_d}{2}\left( \psi-{ \psi}^*\right)\frac{1+ \psi^{2}}{1+| \psi|^{2}} +\partial_{xx}^{2} \psi-\frac{2 \psi^*\left(\partial_x  \psi\right)^{2}}{1+| \psi|^{2}},
\label{LLGStereo}
\end{align}
with $ \psi^*$ being the complex conjugate of $ \psi$. The equilibrium $\mathbf{m}=\mathbf{e_x}$ is mapped to $\psi=0$.

Note that the perpendicular anisotropy coefficient controls both the parametric injection of energy [i.e., terms proportional to $ \psi^*$ in Eq.~(\ref{LLGStereo})] and the cubic saturation terms. Then, we expect that the modulations of $h_d$ will produce appealing dynamical effects. See also that the role of the nonlinear gradients ${2 \psi^*\left(\nabla  \psi\right)^{2}}/({1+| \psi|^{2}})$ in the dynamics can become more critical compared to the one of the relatively week anisotropy-induced saturations.

In the next section, we numerically solve equation~(\ref{Eq-LLG}). However, for pedagogical reasons, let us analyze here its similarities to the pdNLS model. 

Let us consider the scaling $\alpha\sim\vert g\vert\sim\vert h\vert\sim\vert h_d\vert\ll1$, which ensures that each effect (dissipation, detuning, dispersion, and phase-invariance-breaking term) appears once in the equation. This scaling is not difficult to obtain since the current ($g$) and magnetic field ($h$) are control parameters that can be tuned to be of the order of the damping ($\alpha$). Beyond the voltage-controlled magnetic anisotropy~\cite{Suzuki2011,Shiota2009,Kanai2012,Nozaki2012} where $h_d$ is a control parameter but needs an insulating layer, the anisotropy field can be engineered by bulk~\cite{BookSkomski1,BookSkomski2} and interfacial~\cite{Bruno} spin-orbit interactions and magnet thickness, magnetostriction~\cite{Magnetostriction}, periodic modulation of surfaces~\cite{Curvature}, among other mechanisms. With those considerations, Equation~(\ref{LLGStereo}) at linear order, reads
\begin{align}
{\partial_t} \psi=-\mu\psi-i\left(\nu+\partial^{2}_{xx}\right) \psi-i\gamma{ \psi}^*,
\label{EqPDNLSLineal}
\end{align}
where $\mu=-g$, $\nu=-h-{h_d}/{2}$, and $\gamma={h_d}/{2}$. The previous equation is the linear version of the well-known \textit{parametrically driven damped NonLinear Schr\"odinger} (PDNLS) equation, which governs the dynamics of classical systems subject to a force that simultaneously depend on time and the state variable. In those systems, the parameter $\mu$ accounts for the energy dissipation, $\nu$ is the detuning between half the forcing frequency and the natural frequency of the system, and $\gamma$ is the strength of the energy-injection.

Note that in the studied system, the parameters related to the energy injection ($\gamma$) and detuning ($\nu$) depend on the anisotropy field $h_d$. Also, from Eq.~(\ref{EqPDNLSLineal}), we know that patterns or Faraday-type waves~\cite{CoulletPatt} emerge in the magnetic system for $v\geq0$ and $\gamma\geq\mu$~\cite{Equivalence}.

The nonlinear dynamics of nanomagnets with uniform anisotropy coefficients have been extensively studied in the literature. In the next section, we focus our attention on systems with a non-uniform anisotropy. Some mechanisms that can produce a non-uniform anisotropy are films with interfacial anisotropy and variable thickness.

\section{Non-uniform anisotropy field}
\label{NumericalResults}
Let us consider the anisotropy profile
\begin{equation}
h_d(x)=\beta_{zL}e^{-\frac{\left(x-x_L\right)^2}{2\sigma^2}}-\beta_{zR}e^{-\frac{\left(x-x_R\right)^2}{2\sigma^2}},
\label{EqProfileHd}
\end{equation}
where $x_L=L/2-a$ and $x_R=L/2+a$ are the centers of the left and right Gaussians, respectively. The parameter $2a =|x_{L}-x_{R}|$ controls the separation between the Gaussian maxima, $\sigma$ is the characteristic width and $L$ is the device length. Thus, the injection of energy is controlled by four parameters, namely $\{\beta_{zL},\beta_{zR},a,\sigma\}$. When the device is large enough, the spatial average of the Eq.~(\ref{EqProfileHd}) is $\int_0^Ldxh_d(x)\approx\int_{-\infty}^\infty dxh_d(x)=\sqrt{2\pi}\sigma\left(\beta_{zL}-\beta_{zR}\right)$. Then, \textit{no net parametric-like injection of energy occurs for }$\beta_{zL}=\beta_{zR}$. However, rich self-organization is expected from this driving mechanism~\cite{Rabi}. It is worth noting that in other parametrically driven systems with heterogeneous forcing, the qualitative dynamics are insensitive to the specific form of the injection function if its maximum value and characteristic width are similar~\cite{FaradayHeterogeneo}. Furthermore, there is agreement between experiments in fluids vibrated with a square-like spatial profile and theory with Gaussian and parabolic modelings~\cite{FaradayHeterogeneo}.

In addition to the spatial instability mentioned in the previous section, another bifurcation takes place when $\mu$ changes its sign. It occurs when the spin-polarized current injects energy into the system and induces an Andronov-Hopf (oscillatory) instability. At the bifurcation point, $\mu\equiv0$, the oscillator has injection and dissipation of energy only in the nonlinear terms of its equation of motion, and the system has some similarities with the ones with $\mathcal{PT}$-symmetry.

We integrate Eq.~(\ref{Eq-LLG}) using a fifth-order Runge-Kutta algorithm with parameter values similar to the following set: $\alpha = 0.05$, $g     = -0.1$, $h     = -0.4$, $\beta_{zL}   = 1$, $\beta_{zR}   = 1$, $a = 6$, and $\sigma= 3.0$. The simulation box is discretized as $L=dx(N-1)$, with step size $dx=0.25$ and $N=512$ simulation points. We use specular boundary conditions, i.e., the magnetization gradient vanishes at the borders $\left(\partial_x\mathbf{m}\right)\left(0,t\right)=\left(\partial_x\mathbf{m}\right)\left(L,t\right)=0$. The magnetization norm, $\left[\sum_i\mathbf{m}_i^2/N\right]^{1/2}$ is monitored and never deviates from 1 more than $10^{-5}$, which is enough precision for this work.

\subsection{Limit of strong interaction}
\begin{figure}[h]
\includegraphics[scale=.27]{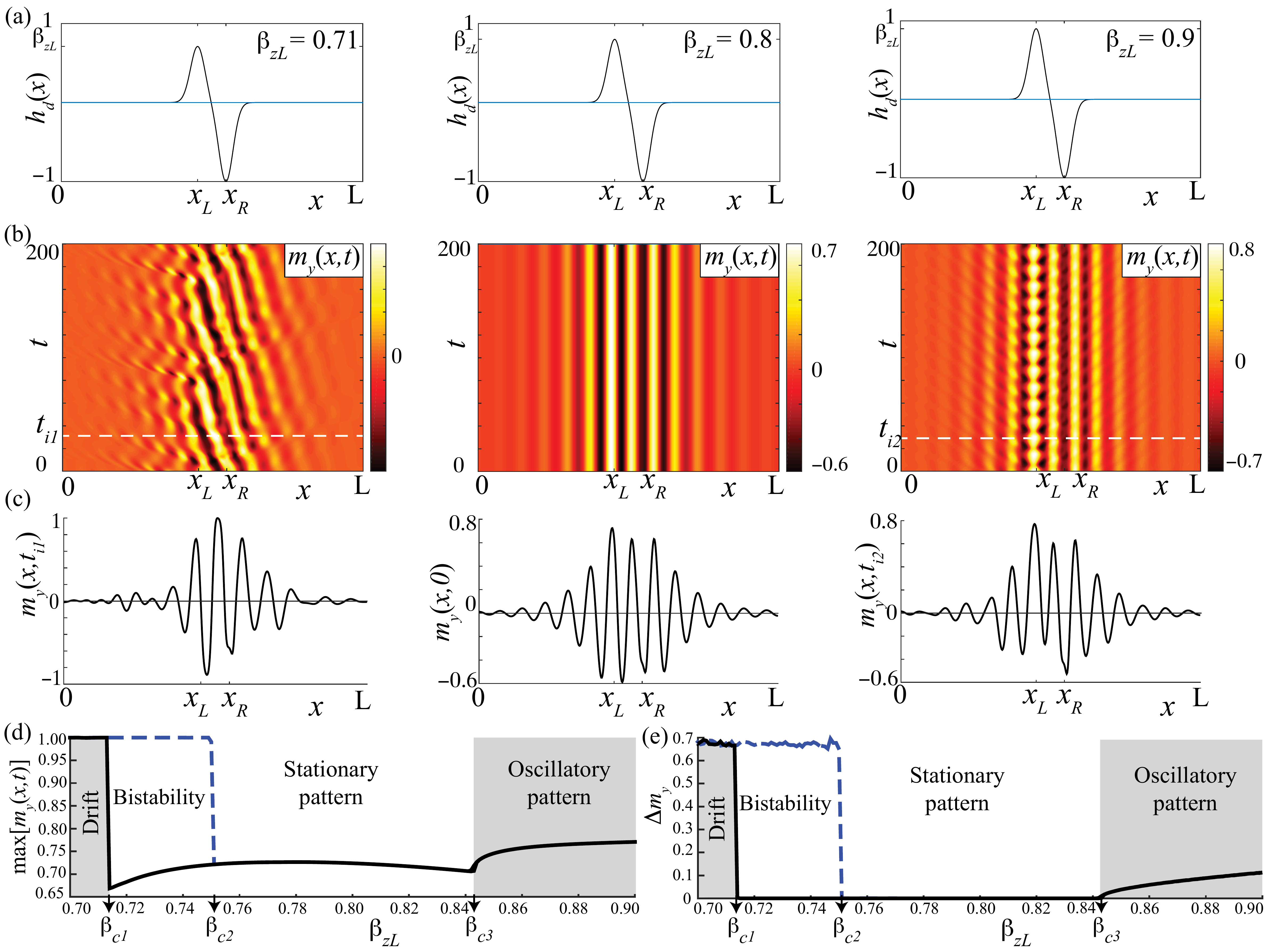}
\caption{Dynamics of \textit{closely interacting} localized patterns. The dynamical regimes of drifting, stationary and phase-oscillatory patterns are exemplified in the left, central, and right panels. (a) shows the anisotropy (or, equivalently, the parametric injection) profile as a function of the position $x$ for $\beta_{zL}=0.71$ (left), $\beta_{zL}=0.8$ (center), and $\beta_{zL}=0.9$ (right). The Gaussian functions are very close, which produces a strong coupling of the localized patterns. Panel (b) is the spatiotemporal diagram of $m_y$ for the same $\beta_{zL}$ values of (a). As these diagrams illustrate, for small $\beta_{zL}$, the patterns have a drifting-like phase dynamics. For moderate values, the localized pattern is stationary. For quasi-symmetric injection parameters, $\beta_{zL}\lesssim\beta_{zR}=1$, a breathing-type phase dynamics appears. (c) shows the snapshots of $m_y$ at the times when the maximum value is reached, max$\left[m_y(x,t)\right]$. (d) Diagram of phases using the maximum value of the magnetization $y-$th component, max$\left[m_y(x,t)\right]$, for a given set of parameters. (e) The oscillation amplitude of the pattern $\Delta m_y$ for a set of parameters. The number $\Delta m_y$ is defined as the standard deviation of $m_y(x_0,t)$, where is the position where the maximum max$\left[m_y(x,t)\right]$ is achieved.}
\label{Fig2}
\end{figure}

Let us start with the limit where the two Gaussians are close to the center but a different amplitude ($\beta_{zL}\neq\beta_{zR}$), see Fig.~\ref{Fig2}(a). As it occurs in parametrically driven systems with homogeneous~\cite{CoulletPatt} and heterogeneous~\cite{FaradayHeterogeneo} energy injections, the parametrically-induced spatial instability creates a texture, as shown in Fig.~\ref{Fig2}(b) and (c) for the spatiotemporal diagrams and snapshots of the $m_y$ variable. We fix $\beta_{zR}=1$ and change the values of $\beta_{zL}$. For
$\beta_{zL}<\beta_{c1}$, the patterns exhibit a drifting-like dynamics where they continuously travel to the center of the simulation. The left-traveling waves have a larger amplitude because they are excited by a stronger forcing $\beta_{zR}>\beta_{zL}$. The left panel of Fig.~\ref{Fig2}(b) and (c) illustrate the drifting solutions. When the $\beta_{zL}$ value is increased above a threshold, stationary patterns emerge subcritically, creating a hysteresis zone in $\beta_{c1}\leq\beta_{zL}\leq\beta_{c2}$. In the region $\beta_{c2}\leq\beta_{zL}\leq\beta_{c3}$, only the stationary patterns are observed, while for $\beta_{zL}>\beta_{c3}$ the pattern undergoes an oscillatory instability that produces breathing-like motions of the pattern phase.
The maximum value of the $m_y$ component, max$\left[m_y(x,t)\right]$, is a single number that measures the amplitude of the pattern. On the other hand, the standard deviation $\Delta m_y$ of the temporal series $m_y(x_0,t)$, where $x_0$ is the position where the max$\left[m_y(x,t)\right]$ is reached, provides useful information of the oscillatory character of the solutions. Figures~\ref{Fig2}(d) and (e) reveal the diagram of phases of the system in the strong interaction limit using max$\left[m_y(x,t)\right]$ and $\Delta m_y$, respectively.

\begin{figure}[t!]
\includegraphics[width=\textwidth]{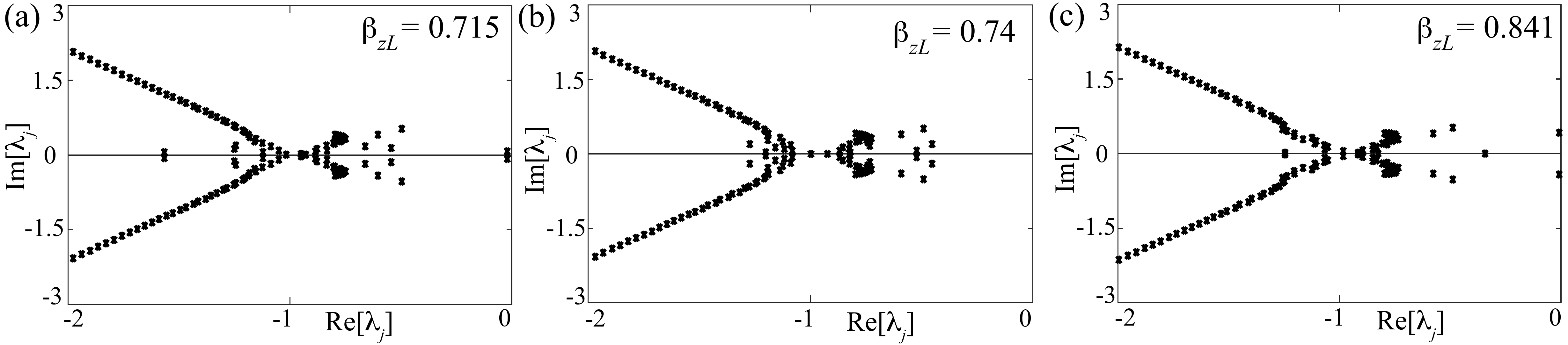}
\caption{Stability spectrum in the \textit{closely interacting} limit. The eigenvalues $\lambda_j$ are plotted in the complex plane, with Re($\lambda_j$) and Im($\lambda_j$) being the real and imaginary parts of $\lambda_j$, respectively. (a) A stationary instability takes place when reducing $\beta_{zL}$. Indeed, when $\beta_{zL}\to\beta_{c1}$, the eigenvalues with the largest real part reduce their imaginary component and cross the vertical axis at the origin. For $\beta_{zL}<\beta_{c1}$, only drifting-like motions are observed, see Fig~\ref{Fig2}. On the other hand, for $\beta_{c1}<\beta_{zL}<\beta_{c2}$, both the stationary and the localized drifting patterns are stable. (b) Spectrum far from any bifurcation bifurcations, all eigenvalues have a negative real part, which characterizes stable states. (c) Onset of the Andronov-Hopf instability. The leading modes have a finite frequency. For $\beta_{c3}<\beta_{zL}$, the localized patterns exhibit a phase-breathing-like oscillatory dynamics, as illustrated on the spatiotemporal diagram of Fig~\ref{Fig2}(b).}
\label{Fig3} 
\end{figure}

The analytic study of the instabilities of non-uniform states in systems with non-uniform parameters is a hard task. However, one can conduct a numerical study as follows. Let us start writing the magnetization vector in the Spherical repsentation, $\mathbf{m}=\sin\theta\left(\cos\phi\mathbf{e_x}+\sin\phi\mathbf{e_y}\right)+\cos\theta\mathbf{e_z}$. The dynamical variables $\theta(x,t)$ and $\phi(x,t)$ are the polar and azimuthal angle, respectively. Integrating Eq.~(\ref{Eq-LLG}), one obtains the stationary solution of the localized pattern, $\left\{\theta_0(x),\phi_0(x)\right\}$, and the small deviations around this state satisfy $\left\{\delta\theta_0,\phi_0\right\}=\sum_je^{\lambda_j t}\vec{u}_j$. The eigenvalues $\lambda_j$ can be obtained numerically after diagonalizing the Jacobian matrix of Eq.~(\ref{Eq-LLG}) for the state $\left\{\theta_0(x),\phi_0(x)\right\}$. Starting from a numerically solved stationary pattern state, Fig.~\ref{Fig3} shows the stability spectrum near the subcritical transition to a drifting pattern in (a), in the zone of stationary patterns in (b), and close to the oscillatory instability in (c). The spectrum allows us to categorize the transitions at $\beta_{zL}=\beta_{c1}$ and $\beta_{zL}=\beta_{c3}$ as stationary and Andronov-Hopf bifurcations, respectively.

\begin{figure}[t!]
\centering \includegraphics[width=0.6\textwidth]{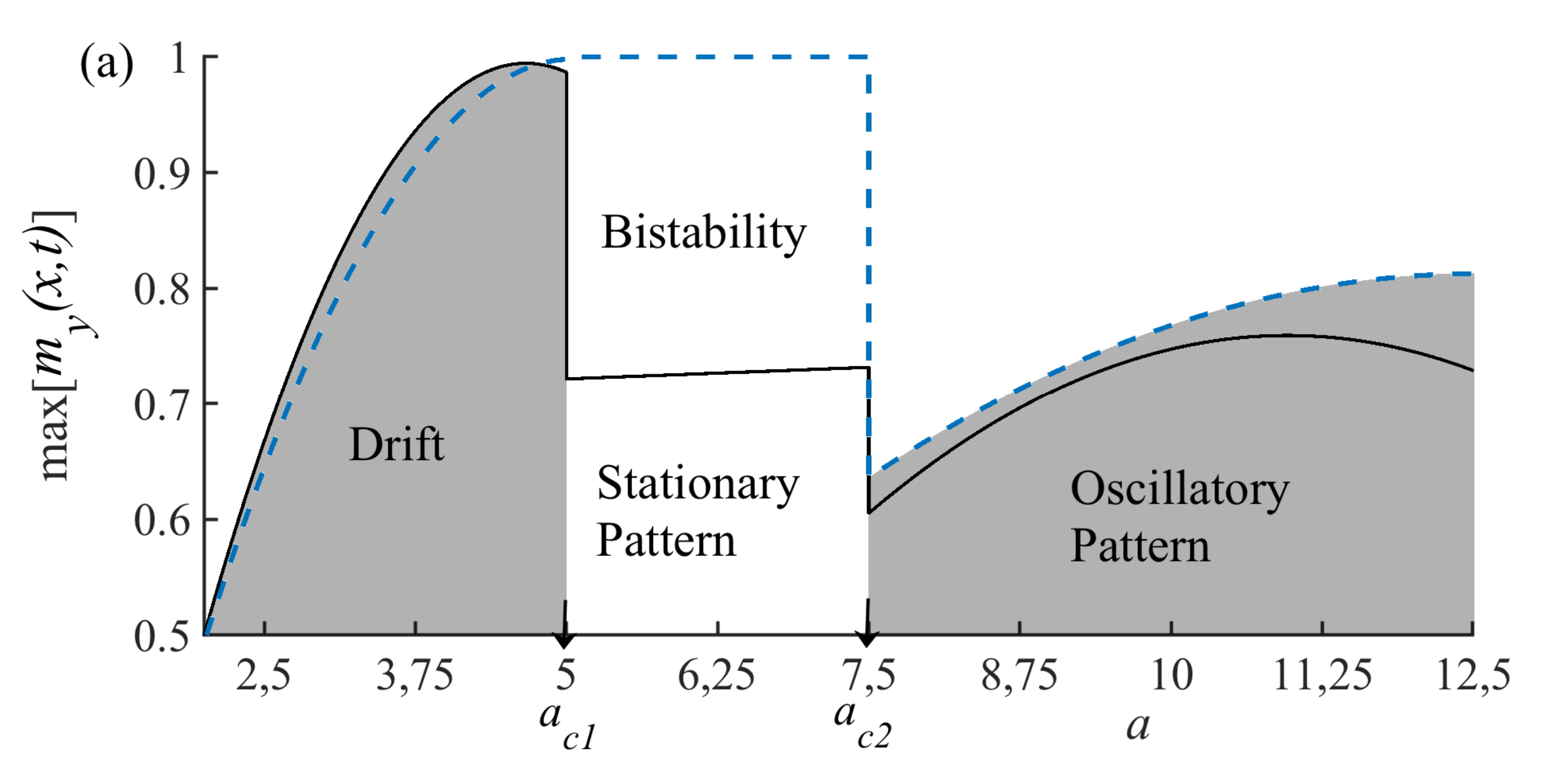}\\
\includegraphics[width=0.32\textwidth]{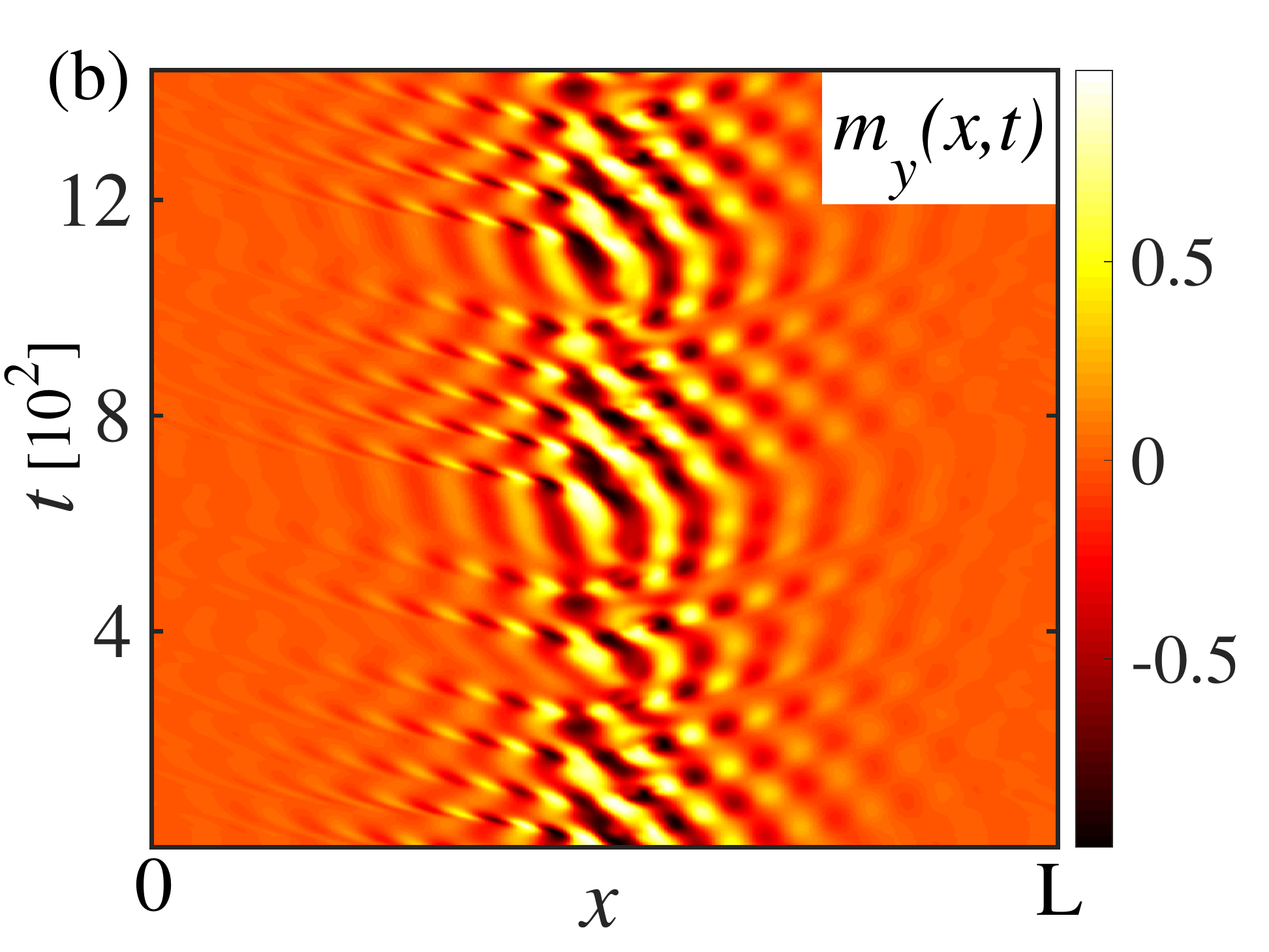}\, 
\includegraphics[width=0.32\textwidth]{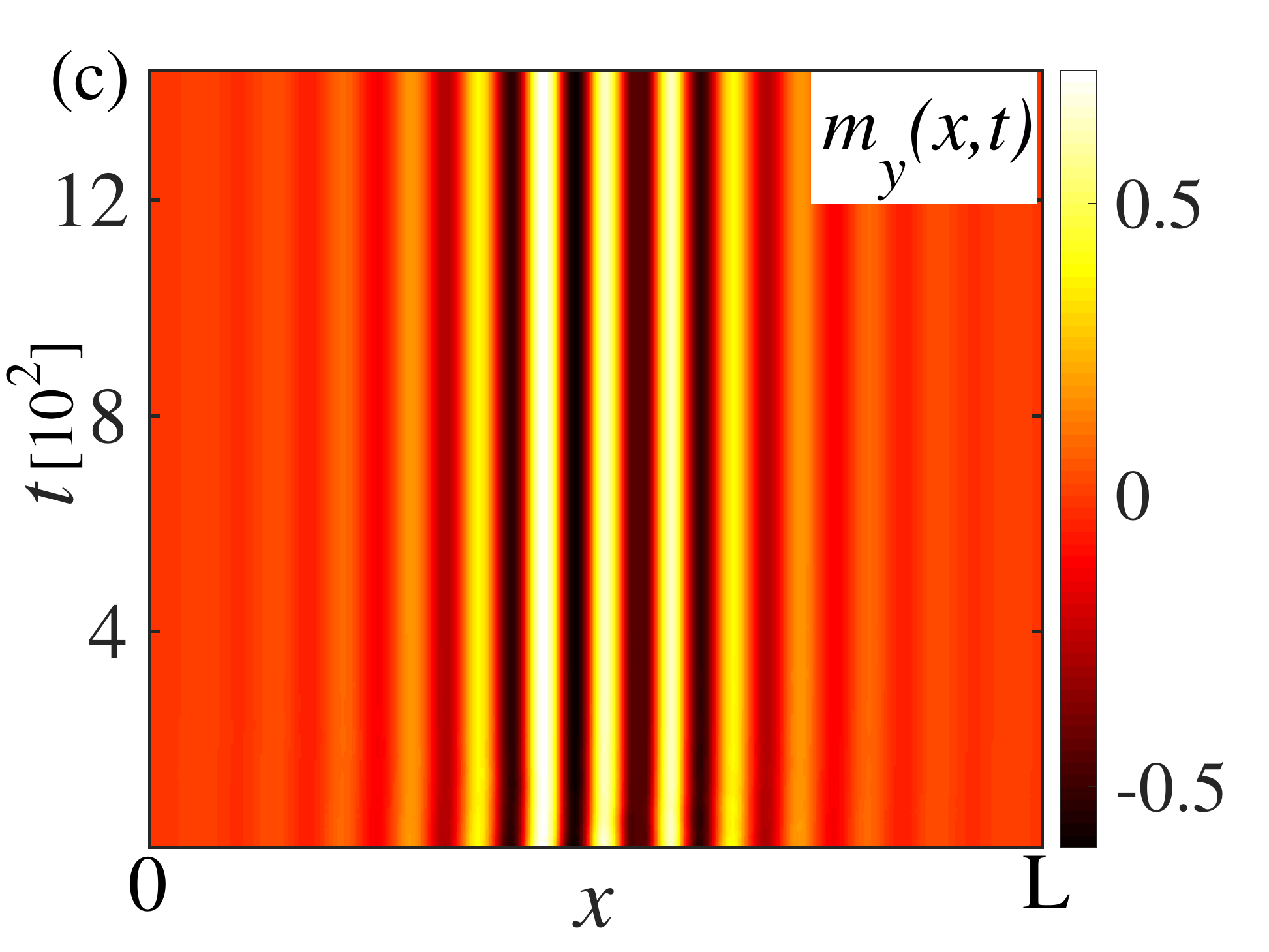}\,
\includegraphics[width=0.32\textwidth]{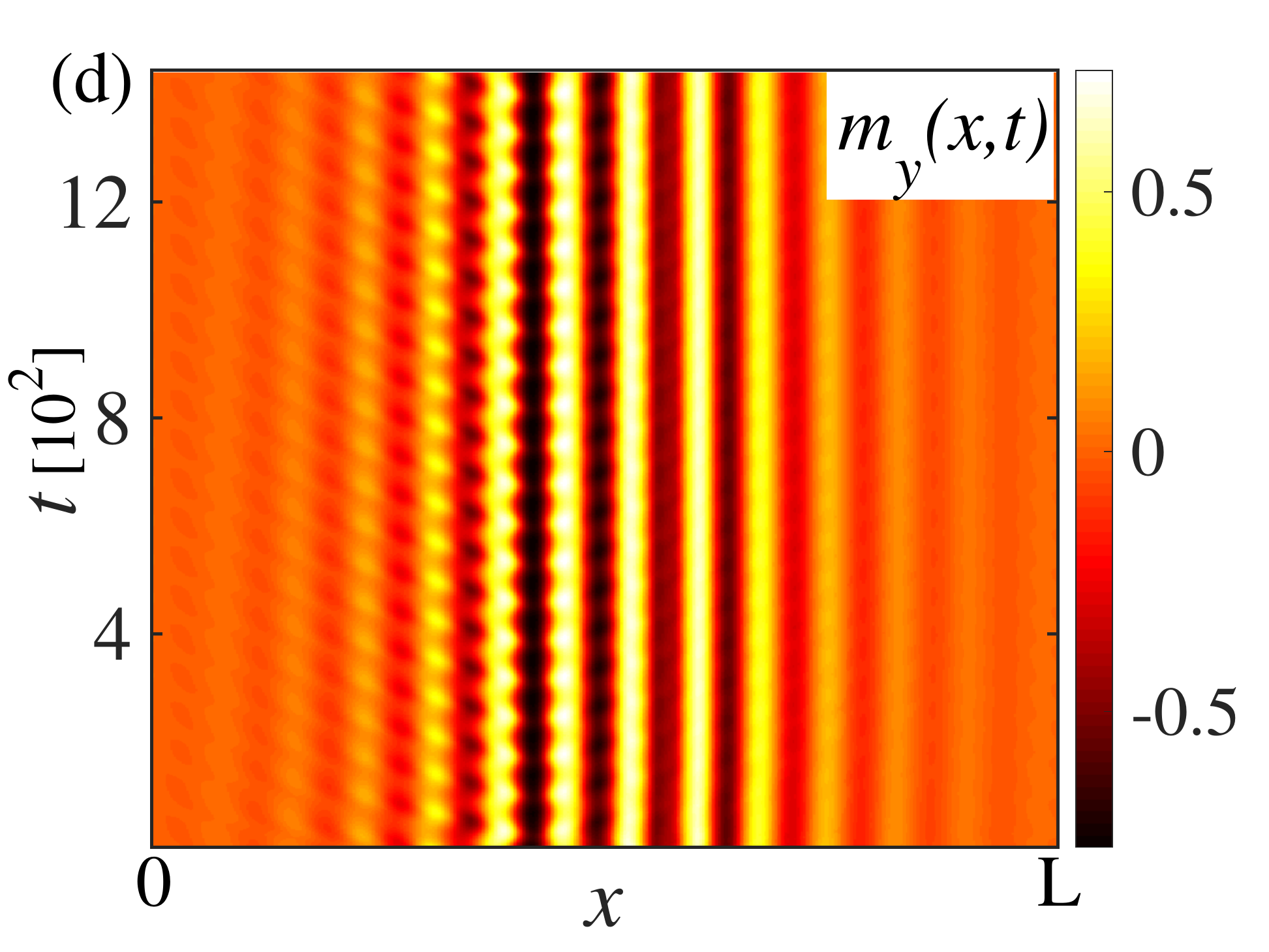}
\caption{Dynamics of interacting localized patterns as a function of the separation parameter $a$. (a) Diagram of phases using the maximum value of the magnetization $y$-th componente, max[$m_{y}(x,t)$] as a function of the separation parameter, for a given set of parameters. For fixed value $\beta_{zL}=0.75$ and $\beta_{zR}= 1$, the patterns have a drifting-like phase dynamics for $a<a_{c1}=5$. For $a_{c1} < a < a_{c2}$ a stationary pattern is observed. A breathing-type phase dynamics appears for $a > a_{c2} = 7,5$.
The dynamical regimes of drifting, stationary and phase-oscillatory patterns are exemplified in (b), (c) y (d), respectively.   }
\label{Fig4}
\end{figure}

Similar results replicate as the separation distance between the left and the right of the anisotropy varies. For $\beta_{zL} = 0.75$ and $\beta_{zR} = 1$ fixed, we varied the separation parameter $a$ in the interval $2\leq a \leq 12.5$. Figure~\ref{Fig4}(a) displays the different dynamical regimes for interacting localized patterns as a function separation parameter $a$. For values smaller than $a_{min} = 2$, the pattern disappears due to the partial cancellation the Gaussians. A drifting-like behaviour is observed for $a_{min}<a<a_{c1}=5$ [shown in Fig.~\ref{Fig4}(b)]. Given both Gaussians are interacting strongly in this limit, the drift is more complex that one observed in the previous case, Fig.~\ref{Fig2}(b). The right-traveling waves struggle to enter into the left region. In contrast, stationary and oscillatory regimes, shown in Fig.~\ref{Fig4}(c) and~\ref{Fig4}(d),  are similar to the ones of Fig.~\ref{Fig2}. The stationary-drifting bifurcation is subcritical-type, showing a bistability region in the interval $a_{c1}<a <a_{c2}= 7.5$, while a supercritical one describes the emergence of oscillatory patterns at the critical value $a_{c2}$. Small differences observed in paths for $a<a_{c1}$ and $a>a_{c2}$ are numerical.
 
\subsection{Diluted limit}
\begin{figure}[t!]
\includegraphics[width=0.9\textwidth]{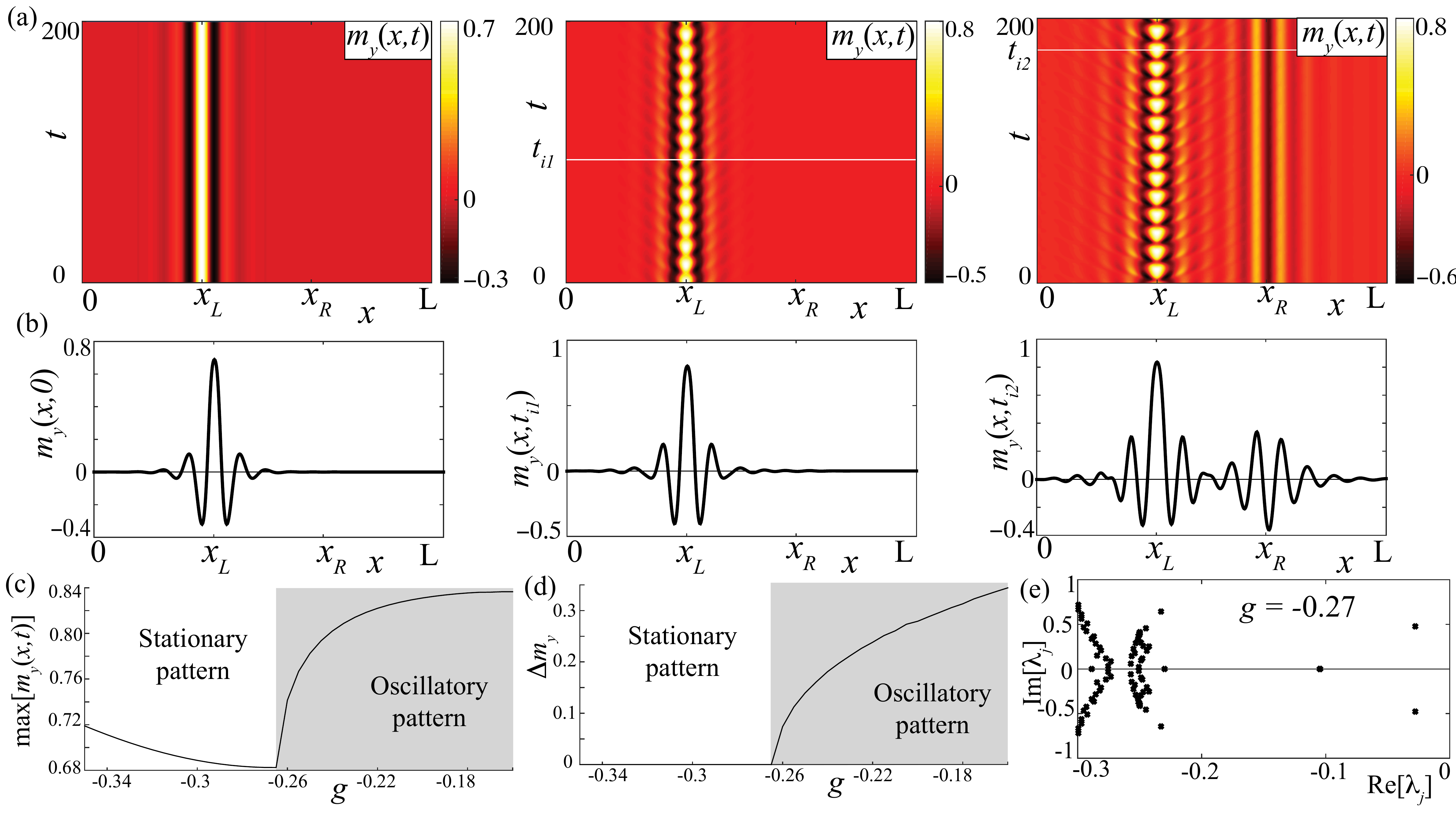}
\caption{Dynamics in the diluted regime. The separation between the anisotropy field peaks is $a=20$, which guarantees that the excited localized states interact only slightly. (a) Spatiotemporal diagram for the stationary localized pattern (left), oscillating texture (center), and both an oscillatory and a static pattern (right). Panel (b) shows a snapshot of the $m_y$ variable at a given time. (c) and (d) show the bifurcation diagram of the left localized pattern in terms of the maximum and standard deviation of the $m_y$ temporal series, equivalent to the graphs of Figs.~\ref{Fig2}(d) and~(e). The stability spectrum is shown in (e), and it reveals a pair of eigenvalues with large imaginary components. As the current parameter $g$ increases, the real part of those eigenvalues becomes positive, leading to the self-sustained oscillations.}
\label{Fig5}
\end{figure}

In this subsection, we present our results in the limit where the Gaussians are far away, and consequently, the dissipative structures at the right and left zones are only slightly interacting. When the injection peaks are well separated, the amplitude mismatch $\beta_{zL}-\beta_{zR}$ is not important for the dynamics. Thus, we use the values $\beta_{zL}=\beta_{zR}=1$ and vary the current parameter $g$. When $g\to0$, the dominant dissipation mechanism vanishes, and the equilibrium $\mathbf{m}=\mathbf{e_x}$ loses its stability. Then, when reducing $\vert g\vert$, we expect the emergence of dynamical states. Figure~\ref{Fig4} summarizes our results. For very negative values of the current $g$, that is for significant dissipation $\mu$, only one pattern emerges, see Fig.~\ref{Fig4}(a) and~(b). When the dissipation goes to zero, the stationary pattern loses stability via a supercritical [see Figs.~\ref{Fig5}(c) and~(d)] Andronov-Hopf [see Figs.~\ref{Fig5}(e)] bifurcation. The new state emits evanescent waves from its core. For even larger values of the current, the right Gaussian stabilizes a static texture.

\section{Conclusions and remarks}
\label{Conclusions}
Nanoscale magnetization dynamics attract considerable attention due to their appealing non-equilibrium behaviors and the associated technological applications. A driving mechanism recently discovered is the modulation of the anisotropy field via the application of voltages, tuning the magnet thickness, and interfacial doping. The temporal modulation of the anisotropy function has gathered considerable attention. However, texture formation and dynamics by heterogeneous anisotropy fields remain mostly unknown. This work is a step in this direction. We studied the self-organization of a one-dimensional magnetic medium driven by an applied magnetic field, a spin-polarized electric current, and a non-uniform anisotropy field. The well-known Landau-Lifshitz-Gilbert-Slonczewski (LLGS) equation describes this system. We concentrated on the parameter sets where the magnet is equivalent to a parametric resonator, even if the magnet is only subject to time-dependent forcing mechanisms. 

The starting point of this manuscript is a simple model based on the one-dimensional and local approximations. However, the numerous and challenging behaviors found justify the use of such approximations as a necessary first step before exploring more elaborate conditions.

The profile of the anisotropy field is a sum of two Gaussian functions with opposite signs. We found a family of localized pattern states. They can be stationary, drifting, or oscillatory textures. In the strong interaction regime, the localized drifting patterns originate from a stationary instability of a static texture, as shown via the calculation of the linear spectrum of the pattern. On the other hand, when the Gaussian functions have similar absolute values, then the patterns undergo an oscillatory (Andronov-Hopf) bifurcation with a breathing phase mode. When the anisotropy field is composed of well-separated peaks, only one transition is observed, namely, localized patterns transit between stationary to phase-oscillatory regimes when the dissipation or injection parameters are modified.

Our findings may motivate the use of localized patterns as effective \textit{individual} oscillations with the capacity to couple and interact, and could potentially be used as units of a nano-oscillators network.

\section*{Acknowledgments}
MAGN thanks to project FONDECYT Regular 1201434 for financial support. FRM-H acknowledges the support of this research by the Dirección de Investigación, Postgrado y Transferencia Tecnológica de la Universidad de Tarapacá, Arica, Chile, under Project N$^{\circ}$ 4742-20. AOL acknowledges financial support from Postdoctorado FONDECYT 2019 Folio 3190030, and Financiamento Basal para Centros Cientificos de Excelencia FB08001.

%%%%%%%%%%%%%%%%%%%%%%%%%%%%%%%%%%%%%%%%%%%%%%%%%%%%%%%%%%%%%%%%%%%%%%%%%%%%%%%%%%%%%%%%%%%%%%%%%%%%%%%%%%%%%%%%%%%%%%%%%%

\end{document}